\documentclass[twocolumn,showpacs,preprintnumbers,amsmath,amssymb,prl,floatfix]{revtex4}

\usepackage{graphicx}
\usepackage{dcolumn}
\usepackage{bm}
\usepackage{longtable}
\usepackage{epsfig}
\usepackage{times}

\begin{document}

\title{Isomer triggering via nuclear excitation by electron capture}

\author{Adriana \surname{P\'alffy}}
\email{Palffy@mpi-hd.mpg.de}

\author{J\"org \surname{Evers}}
\email{Evers@mpi-hd.mpg.de}

\author{Christoph~H. \surname{Keitel}}
\email{Keitel@mpi-hd.mpg.de}

\affiliation{Max-Planck-Institut f\"ur Kernphysik, Saupfercheckweg~1, 
69117 Heidelberg, Germany}

\date{\today}

\begin{abstract}Triggering of long-lived nuclear isomeric states via coupling to the
atomic shells in the process of nuclear excitation by
electron capture (NEEC) is studied. NEEC occurring in highly-charged ions
can excite the isomeric state to a triggering level that subsequently
decays to the ground state. We present total cross sections for NEEC
isomer triggering considering experimentally confirmed low-lying
triggering levels and reaction rates based on realistic
experimental parameters in ion storage rings. A comparison with  other isomer
triggering mechanisms shows that, among these, NEEC is the most efficient.

\end{abstract}


\maketitle
\begin{figure}
\begin{center}
\includegraphics[width=0.4\textwidth]{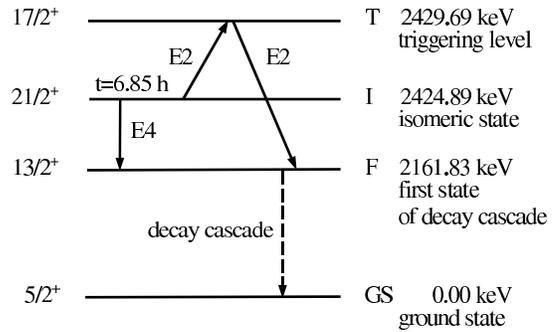}
\caption{\label{mo93} Partial level scheme of $^{93}_{42}\mathrm{Mo}$. As a general notation, the isomeric state ($I$) can be excited to the triggering level ($T$). The triggering level subsequently decays back to $I$ or to a level $F$ which initiates a cascade via different intermediate states (dashed line) to the ground state ($GS$). $E4$ denotes a strongly hindered nuclear ($\gamma$) transition.}
\end{center}
\end{figure}

Metastable nuclear states (nuclear isomers) have been the subject of intense debate in the last years, not least fueled
by a series of controversial experimental evidence \cite{Collins,Ahmad}. Isomeric triggering---release on demand of the energy stored in the excited metastable nuclear state---has been proposed via a number of nuclear excitation mechanisms, such as photoabsorption \cite{Kara_Carr1,Carr_Kara}, Coulomb excitation \cite{Hayes,Kara_Carr5} or coupling to the atomic shells \cite{Kara_Carr,Tkalya,Karpeshin}. 
The isomeric state could be thus excited to a higher level which is associated with freely radiating states and therefore releases the energy of the metastable state, as schematically presented in Figure~\ref{mo93} for the case of $^{93\rm{m}}$Mo. So far, attempts to trigger the energy release from isomers were focused on the 31-year  $^{178\rm{m2}}{\rm Hf}$  isomer, which became  a highly controversial issue \cite{Collins,Ahmad,Carroll}. As an undisputed result stands the triggering of the $^{180\rm{m}}{\rm Ta}$ quasistable isomer, which has a lifetime of $t_{1/2}\ge 1.2\times 10^{15}$~yr, longer than the expected age of the universe, by  2.8~MeV bremsstrahlung photons \cite{Belic,Phil}. 
The motivation for the study of isomers is mainly twofold~\cite{walker,nat_phys}. First, a better understanding of the properties of nuclear isomers, their formation and their decay mechanisms is desirable as such, for example, because of the 
role isomers play in the creation of the elements in the universe. But much interest also arises from a number
of fascinating potential applications related to the controlled release of nuclear energy on demand, such as nuclear batteries.
For both purposes, efficient triggering mechanisms need to be identified. Motivated by this, here we draw  attention  to triggering via nuclear excitation by electron capture (NEEC). In the context of isomer research, unlike other
excitation mechanisms, NEEC  has received little attention until now. In Ref.~\cite{Zader}, limited access to important internal conversion coefficient values lead to qualitative estimates off
by many orders of magnitude. NEEC has also been  mentioned as possible triggering mechanism in the study of enhanced nuclear decay in astrophysical hot dense plasmas \cite{Morel},  and as nuclear excitation mechanism of the 
$^{235\rm{m}}\mathrm{U}$ isomer in laser generated plasmas \cite{Harston}. 

In this Letter, we 
discuss in detail  the feasibility of NEEC-assisted isomer activation, and derive triggering probabilities using
a rigorous treatment of the electron-nucleus interaction for a number of isomers of interest. 
As our main result, we find that typically NEEC is most efficient among the investigated isomer triggering mechanisms.

 In the resonant process of NEEC, a free electron is captured into a highly-charged ion with the simultaneous excitation of the nucleus \cite{neec}. NEEC is the time-reversed process of internal conversion (IC) and is related to other processes that couple nuclear transitions to the electronic shell such as nuclear excitation by electron transition (NEET) and bound internal conversion (BIC). 
While for broadband photoexcitation of isomers with bremsstrahlung photons knowing the existence or exact position of the triggering level is not necessary, for resonant processes such as NEEC or NEET this is a more sensitive matter. Levels above the isomer are often not connected by transitions to the metastable state, as they belong to different nuclear bands.  It is therefore unlikely  that such a hindered transition could be triggered in an experiment either by photons or by coupling to the atomic shell.  We envisage NEEC isomer triggering via the nearest triggering level lying above the metastable state that has been either experimentally confirmed or for which the corresponding transition is theoretically predicted. Low-lying triggering levels are desirable for obtaining high energy gain and facilitating the excitation to the triggering level. 
Characteristic data for few interesting cases is summarized in Table~\ref{nds_data}.

In the case of the $^{235}_{92}\mathrm{U}$ isotope there is no low-lying triggering level that connects the 76~eV isomeric level with the ground state, but the excitation of the isomeric state to a  triggering level at 51.701~keV that will in turn decay back to the isomeric state via an intermediate level is possible. While one would not gain the energy of the isomeric level (which in this case is only 76~eV), this isotope offers the possibility of testing the excitation mechanism at very low energy by detecting the photons emitted in the degenerate triggering.

Triggering of the 48.6~keV isomeric state of  $^{242\rm{m}}{\rm Am}$ has been previously investigated \cite{Zader}. The long-lived $^{242\rm{m}}{\rm Am}$ isomer ($t_{1/2}\sim 141$~years) can be produced by neutron capture with a significant cross section \cite{Carroll} and has an instable ground state with a 16~h lifetime.  The level data in \cite{Am242} show that at just 4.1~keV above the $5^-$ isomer lies a $3^-$ level that could be reached by a $E2$ transition, which however hasn't been observed experimentally. This lack of experimental observation could be due to a $K$ hindrance or the small transition energy. The triggering of the isomeric state via NEEC would in turn not only release the 48.6~keV stored by the isomer, but would also begin the transmutation chain of the ground state. The authors of Ref.~\cite{Zader} have been comparing several mechanisms of non-radiative triggering of $^{242\rm{m}}{\rm Am}$, including NEEC. In their qualitative estimate of the NEEC triggering cross section, however, a key value in the calculation---the partial IC coefficient---was reported to be unknown.
This coefficient is the ratio between the IC and radiative decay rates for a certain nuclear transition 
involving the electrons of a particular atomic orbital or shell. 
As stated in Ref.~\cite{Zader}, this coefficient was assumed somewhat arbitrarily in order to numerically
estimate a cross section. This led to their conclusion that NEEC triggering is less favorable than
other triggering mechanisms.
In the following, we derive the partial IC coefficient using a rigorous treatment of the electron-nucleus interaction, and obtain results that are many orders of magnitude above the ones in Ref.~\cite{Zader}.

We calculate the total cross section $\sigma$ for NEEC to the triggering level followed by its subsequently decay via a branching ratio releasing the energy of the isomeric state. The initial decay of the triggering level $T$ to the $F$ level (see Fig.~\ref{mo93}) occurs either by emitting a photon or by IC. The electron capture is considered to occur in bare ions, with the exception of $\rm{Am}$, where we have considered as initial state the $\rm{Am}^{83+}$ ion, with the $K$ and $L$ shells and the $3s$ orbital occupied. The capture orbitals are presented in Table \ref{res_str}. If the nuclear transition energy does not allow for the capture into the $K$ shell and recombination occurs into higher shells, the subsequent fast electronic decay to the ground state is taken into account. The IC decay channel of the triggering level $T$ back to the isomeric state $I$ is therefore completely changed. For the case of heavy highly-charged ions like uranium or lead this can be of great advantage. In these cases, we consider capture into the $L$ shell. The captured electron will decay rapidly to the ground state, therefore inhibiting completely the IC decay channel of the triggering level $T$ back to the isomeric state.

\begin{table}
\caption{\label{nds_data} Considered isomeric states and triggering levels.
The lifetimes and energies of the isomeric states are given in the second and third column, respectively. The energy of 
the triggering level $E_T$ and the multipolarity $L$  of the transition $I\!\to \!T$ are presented in the last columns. 
Data taken from \cite{nds}.}
\begin{ruledtabular}
\begin{tabular}{lcccc}
$^A_Z\mathrm{X}$ & $t_{1/2}$  & $E_{\rm{m}}$ (keV) & $E_{T}$ (keV) & $L$ \\ \hline

$^{93}_{42}\mathrm{Mo}$   & 6.85 h & 2424.89 & 2429.69 & $E2$\\
$^{152}_{63}\mathrm{Eu}$   & 9.31 h & 45.599 & 65.296& $M1$\\
$^{178}_{72}\mathrm{Hf}$   & 31 yr & 2446.05 & 2573.5  & $M2$\\
 $^{189}_{76}\mathrm{Os}$  & 5.8 h& 30.812 & 216.661 & $M1+E2$\\
$^{204}_{92}\mathrm{Pb}$  & 67.2 m & 2185.79 & 2264.33 & $E2$\\
$^{235}_{92}\mathrm{U}$   &  26 m & 0.076 & 51.709  & $E2$ \\
$^{242}_{95}\mathrm{Am}$   &  141 yr & 48.60 & 52.70  & $E2$ \\

\end{tabular}
\end{ruledtabular}
\end{table}

The total resonance strength, i.e., the integral of the total cross section $\sigma$ over the continuum electron energy, for NEEC followed by the decay of the triggering level to the state $F$ is given by \cite{neec}
\begin{equation}  
S^{I\!\to\! F}_{\rm NEEC}=\frac{2\pi^2}{p^2} Y^{I\!\to\! T} B^{T\!\to\! F}  \, ,
\label{resstr}
\end{equation}
where $p$ denotes the momentum of the continuum electron,  $Y^{I\!\to\! T}$ is the NEEC rate and $B^{T\!\to\! F}$ is the branching ratio for the decay of the triggering level $T$ avoiding the isomeric state. The branching ratio is given by the ratio between the radiative and IC decay rates from the triggering level $T$ to the $F$ state and the total width of the $T$ state, $B^{T\!\to\! F}=(A_{rad}^{T\!\to\! F}+A_{IC}^{T\!\to\! F})/\Gamma_{T}$. We adapt the formalism presented in \cite{neec} to account for the more complicated nuclear level structure close to the isomeric state. The electron-nucleus interaction occurs either by Coulomb interaction or by  virtual photon exchange between the electronic and nuclear currents. The nucleus is described using a collective model \cite{Ring} which allows expressing the nuclear matrix element with the help of the reduced transition probabilities, for which we consider experimental values.  Nuclear data  were taken from \cite{nds}. Where the experimental reduced transition probabilities were not available, single-particle (Weisskopf) estimations \cite{Ring} have been used. The electronic motion is considered relativistically and taking into account the finite size of the nucleus.
 The electronic energy levels and the radial wavefunctions for the corresponding orbitals are calculated with the GRASP92 package \cite{grasp92}. The IC rates are derived from the corresponding NEEC rates using the principle of detailed balance in order to account for the real number of electrons available for IC in the ion. 
The values for the NEEC resonance strength for isomer triggering are presented in Table~\ref{res_str}. As a comparison, photoexcitation resonance strengths using the same triggering levels are also presented. 
No NEET values are given, since among all the considered isotopes, $^{242}_{95}\mathrm{Am}$ is the only case 
for which the match between electronic and nuclear energies allows NEET, as will be discussed later. 
The  photoexcitation resonance strength $S^{I\!\to\! F}_{\rm{x\,ray}}$ has an expression  equivalent to the one in Eq.~(\ref{resstr}), involving the photon momentum and nuclear photoexcitation rate. The latter can be related with the help of the detailed balance principle to  radiative decay rates. Note that the branching ratios in photoexcitation and NEEC triggering
can be different depending on the initial electronic configuration of the isomer. 

\begin{table}
\caption{\label{res_str} Total resonance strengths $S$ for NEEC and x-ray triggering of isomers. NEEC occurs in the $nl_j$ orbital. The continuum electron energy at the resonance is denoted by $E_c$. }
\begin{ruledtabular}
\begin{tabular}{lcrcc}
$^A_Z\mathrm{X}$ & $nl_j$  & $E_c$ (keV) & $S^{I\!\to\! F}_{\rm{NEEC}}$ (b eV) & $S^{I\!\to\! F}_{\rm{x\,ray}}$ (b eV) \\ \hline

$^{93}_{42}\mathrm{Mo}$   & $3p_{3/2}$ & 2.113 & $9.1\times 10^{-6}$ & $1.4\times 10^{-8}$\\
$^{152}_{63}\mathrm{Eu}$   & $2s_{1/2}$ & 5.204 & $3.4\times 10^{-4}$& $6.5\times 10^{-5}$\\
$^{178}_{72}\mathrm{Hf}$   &$1s_{1/2}$  & 51.373 &  $2.0\times 10^{-7}$ &$5.4\times 10^{-8}$ \\
 $^{189}_{76}\mathrm{Os}$  &$1s_{1/2}$ & 131.050 & $1.2\times 10^{-3}$ & $2.2\times 10^{-2}$\\
$^{204}_{92}\mathrm{Pb}$  & $2p_{3/2}$ & 55.138 & $4.9\times 10^{-5}$ & $8.7\times 10^{-6}$\\
$^{235}_{92}\mathrm{U}$   & $2p_{1/2}$  & 21.992 &  $1.3\times 10^{-1}$& $1.3\times 10^{-2}$ \\
$^{242}_{95}\mathrm{Am}$   &$5p_{3/2}$  & 0.135 & $3.6\times 10^{-3}$  & $2.4\times 10^{-8}$ \\

\end{tabular}
\end{ruledtabular}
\end{table}

For the case of isomer triggering of $^{242}_{95}\mathrm{Am}$ electron capture into the $5p$ orbital of the $\rm{Am}^{83+}$ ion has been considered. Due to the very low nuclear excitation energy of only 4.1~keV, capture into the $5p$ orbital of the bare ion is not possible. The $5p$ orbital is particularly interesting as it has the largest NEEC rate values among the orbitals of the $O$ shell. The partial IC coefficient that we have obtained for the  $5p$ orbital is $\alpha_{5p}=6\times 10^5$, more than ten orders of magnitude larger than the one assumed in Ref.~\cite{Zader}. 
Our value is consistent with the IC coefficients for $Z=95$ and transition energy 4.1~keV tabulated in Ref.~\cite{ICC_tabulation}.
While our estimated value of the x-ray triggering resonance strength agrees with the x-ray cross section presented in \cite{Zader}, our NEEC resonance strength for the triggering of  $^{242\rm{m}}{\rm Am}$ is 6 orders of magnitude larger than the x-ray photoexcitation one.  A further comparison with the NEET and Coulomb excitation cross section estimations for $^{242}_{95}\mathrm{Am}$ presented in Ref.~\cite{Zader} proves that NEEC is the most efficient triggering mechanism for this isomer.

The values in Table~\ref{res_str} show that  for the considered  low-lying triggering levels, the NEEC nuclear excitation mechanism is more efficient than the photoabsorbtion one, with the exception of $^{189}_{76}\mathrm{Os}$. 
For this isomer, the photoexcitation rate is higher than the NEEC rate because of the large transition energy (185~keV).
While the radiative rate for $M1$ transitions is proportional to  $E^3$, where $E$ denotes the nuclear transition energy, the NEEC resonance strength is in turn proportional to the inverse of the squared continuum electron momentum and decreases for high energies. 
As a general rule, the NEEC rate is larger for isotopes with higher atomic number $Z$. For example, at similar transition energies and multipolarities, the NEEC rate for isomeric triggering in $^{242}_{95}\mathrm{Am}$ is almost two orders of magnitude larger than the one in $^{93}_{42}\mathrm{Mo}$.
 On the other hand, very low nuclear transition energies are usually associated with small photoexcitation or NEEC rates and the corresponding decay rates. For $^{93}_{42}\mathrm{Mo}$, the resonance strength value of $9.1\times 10^{-6}$ is related to the small NEEC rate. 
The largest resonance strength in Table~\ref{res_str} is the one for the degenerate triggering of the 76~eV isomeric state of  $^{235}_{92}\mathrm{U}$. In this case, a rather large NEEC rate for the capture of the electron in the $L$ shell of a bare ion is corroborated with an efficient branching ratio, related to the rearrangement of the electronic configuration. The captured electron performs a fast decay to the $K$ shell, and therefore completely inhibits IC decay of the triggering level back to the isomeric state.

While NEET and BIC have been both confirmed experimentally \cite{Kishimoto,Carreyre}, the only performed experiment aimed at NEEC observation so far at GANIL \cite{ganil} did not succeed. 
Experimental observation of NEEC and furthermore NEEC isomeric triggering  is certainly a demanding task.
Nevertheless, with new ion storage ring facilities~\cite{fair} and numerous powerful electron beam ion traps around the world,  a successful NEEC experiment is now much closer to reality. 
 A difficulty that all nuclear excitation mechanism share comes from the very narrow natural widths of the nuclear excites states, requiring an energy match with high accuracy. In comparison, however, NEEC has at least the advantage of more flexibility in choice of isotope and   envisaged transition, since energy conservation is fulfilled by a continuum electron. In the experimental scenarios envisaged up to now, in which the nuclear excitation would occur for the nucleus in the ground state, radiative recombination (RR) is the dominant electron recombination channel and  is responsible for interference effects and a strong background \cite{interference}. 
An important new 
aspect when considering isomer triggering via NEEC is that the energies of the excitation from the isomeric state and of 
the photons emitted in the decay of the triggering level are different.
Thus the experimental background due to RR is much reduced due to the different photon energies in the involved
processes.  In this sense, our concept of isomer triggering by NEEC is also an interesting candidate
for an experimental verification of NEEC by itself.

The new Synchrotron SIS100 that will be built in the future at FAIR \cite{fair} will provide for storage ring experiments a beam of $10^{11}$ ions, delivered in a 50~ns long bunch \cite{Tahir}. 
Isomeric beams produced in nuclear reactions and subsequently selected through the fragment separator Super-FRS are expected to have a lower intensity, of up to $10^6$ particles per bunch \cite{frs}. The ions can be stored into the experimental storage ring  and cycle through a dense electron target provided by a liquid helium beam \cite{Grisenti}. Depending on the actual experiment, the lifetime of the ion beam in the storage ring determines the beam reinjection frequency  from the synchrotron.  We can estimate the reaction rate for isomeric triggering according to the expression \cite{Currell}
\begin{equation}
R=\int dV\: n_i\,n_e\:  \langle \sigma v \rangle \:  \frac{\Gamma_n}{\Gamma_e}\, ,
\end{equation}
where $dV$ is the interaction volume, $n_i$ and $n_e$ represent the ion and electron density, respectively, $\sigma$ is the NEEC triggering cross section, whose integrated values are given in Table~\ref{res_str}, $v$ the relative velocity of the ions with respect to the electrons and the ratio $\Gamma_n/\Gamma_e$ gives the extent to which the energy resonance condition is fulfilled. The relative velocity in the ion-electron collisions is mainly given by the ions cycling in the storage ring at the resonance energy required for NEEC. 
The  electron density of the liquid beam electron target is $10^{15}$~cm$^{-2}$ \cite{Grisenti}, with an energy spread given by the Compton profile of around 1~keV. With an estimate of $10^6$ ions in the isomeric beam \cite{frs}, we obtain a  reaction rate of $6.5\times 10^{-2}$~s$^{-1}$ for the NEEC triggering of $^{235\rm{m}}{\rm U}$ and of $1.1\times 10^{-3}$~s$^{-1}$ for the case of $^{189\rm{m}}{\rm Os}$. The NEEC triggering reaction rate for the case of $^{178\rm{m2}}{\rm Hf}$ is unfortunately much lower, just $1.5\times 10^{-7}$~s$^{-1}$, due to the high multipolarity ($M2$) of the triggering transition and the high nuclear transition energy of 127~keV to the  triggering level. The bremsstrahlung photoabsorbtion experiments with $^{178\rm{m2}}{\rm Hf}$ described in Refs.~\cite{Collins,Ahmad} were investigating the possibility of a $K$-mixing level at lower energies, initially at 40~keV and later at around 10~keV. While in \cite{Collins} a $K$-mixing level was identified that would trigger the isomer decay at 2457.20~keV,  this result could not be confirmed by other experimental groups. Since for NEEC triggering the knowledge of the triggering level energy of the occurring transitions is necessary, it is inefficient to use this mechanism for a search of possible triggering levels. However, once a triggering level is identified by other methods, NEEC triggering can be used as an energy-selective consistency check.

In conclusion, our theoretical calculations show that coupling to the atomic shells via NEEC can be the most efficient isomer triggering mechanism when low-lying triggering levels are involved. The different nuclear transition energies for the excitation and deexcitation of the triggering level make isomers good candidates for the experimental observation of NEEC. Considering the high electron densities in hot astrophysical plasmas, NEEC isomer triggering calculations and experiments may be relevant beyond laboratory experiments.

\vspace{0.3cm}

The authors would like to thank Zolt\'an Harman and Stefan Schippers for  fruitful discussions.


\end{document}